\documentclass[conference,10pt]{IEEEtran}

\usepackage[utf8]{inputenc}
\usepackage[T1]{fontenc}
\usepackage{cite}
\usepackage{amsmath,amssymb,amsfonts}
\usepackage{algorithmic}
\usepackage{graphicx}
\usepackage{textcomp}
\usepackage{xcolor}
\usepackage{booktabs}
\usepackage{multirow}
\usepackage{hyperref}
\usepackage{url}
\usepackage{siunitx}
\usepackage{tikz}
\usetikzlibrary{arrows.meta,positioning,shapes.geometric,fit,backgrounds}
\hypersetup{colorlinks=true,linkcolor=blue,citecolor=blue,urlcolor=blue}

\newcommand{\fastgrnn}{FastGRNN}
\newcommand{\msp}{MSP430G2553}
\newcommand{\arduino}{Arduino Uno R3}
\newcommand{\hapt}{HAPT}

\title{From Compression to Deployment:\\
Real-Time and Energy-Efficient \fastgrnn{}\\
on Ultra-Constrained Microcontrollers}

\author{
\IEEEauthorblockN{Emre Can Kızılateş}
\IEEEauthorblockA{Electrical and Electronics Engineer\\
Independent Researcher, Izmir, Turkey\\
kizilatesemrecan@gmail.com}
}

\begin{document}

\maketitle

\begin{abstract}
The dominant trajectory of modern machine learning has been to
scale \emph{up}: larger models, larger accelerators, larger memory
budgets. Yet a multi-year global semiconductor supply
constraint~\cite{burkacky2022chipshortage,khan2021chipshortage} and
the growing energy and carbon cost of always-online
inference~\cite{strubell2019energy,patterson2021carbon} expose the
fragility of this trajectory and motivate the opposite direction:
\emph{refactoring} AI and ML algorithms to fit the small,
ubiquitous microcontrollers that are already in mass production in
wearables, sensors, and edge appliances. This
paper is a study in that direction. We present an end-to-end
open-source reproduction of
\fastgrnn{}~\cite{kusupati2018fastgrnn}, a compact gated recurrent
cell, deployed on two bare-metal targets representative of the
low end of the available silicon supply: the 8-bit \arduino{}
(ATmega328P) and the 16-bit \msp{} (no hardware multiplier;
16~KB Flash; 512~B SRAM). Our compression pipeline combines
low-rank weight factorization, iterative hard-thresholding
sparsity, and per-tensor Q15 post-training quantization with
explicit activation calibration. The deployed model occupies \textbf{566~bytes of weights} and
achieves \textbf{macro F1 $=$ 0.918} (seed~0; five-seed Q15
mean $0.853{\pm}0.107$) on the \hapt{} test set.
It matches a PyTorch reference at \textbf{100\% prediction
agreement} across 3{,}399 test windows (MCU seed~0; 99.91--100\%
C-equivalent across five seeds).
Both platforms sustain \textbf{real-time 50~Hz streaming
inference} (9.21~ms per sample on Arduino; 13~ms on MSP430), where
a 256-entry sigmoid/tanh look-up table delivers a
\textbf{30.5$\times$ speedup} on the multiplier-less MSP430.
Four contributions extend the original \fastgrnn{} paper:
(i) cross-platform bit-equivalent deterministic inference;
(ii) characterization of recurrent warm-up latency (median 74 samples,
1.48~s; worst-case 125 samples, 2.50~s over 100 test windows) that
bounds end-to-end response time;
(iii) a deployable look-up-table recipe for multiplier-less embedded
targets delivering a \textbf{30.5$\times$} end-to-end speedup; and
(iv) a hardware energy characterization showing \textbf{17.7~mW}
active inference power, \textbf{$<$0.09~mW} idle power, and a
\textbf{96.7\%} reduction in energy per inference window with the LUT.
Together, these results demonstrate that compact recurrent
architectures, combined with calibrated quantization, look-up-table
activations, and measured energy profiling, can deliver accurate,
memory-efficient, and energy-efficient human activity recognition on
ultra-resource-constrained microcontrollers without specialized
accelerators.
\end{abstract}


\section{Introduction}
\label{sec:intro}

\subsection{Motivation: Scaling Down When Silicon Will Not Scale Up}

The last decade of machine learning has been characterized by a
clear and consistent assumption: that the appropriate response to a
hard problem is a larger model trained on a larger accelerator with
a larger memory budget. This trajectory has produced spectacular
results in language, vision, and multimodal
generation~\cite{patterson2021carbon}. It has also imposed energy,
carbon, and supply-chain costs that are now visible at a
macroeconomic scale: a multi-year global shortage of leading-edge
semiconductors disrupted the automotive, consumer electronics, and
medical-device industries between 2020 and
2024~\cite{burkacky2022chipshortage,khan2021chipshortage}, and the
energy footprint of always-online inference is large and growing
faster than supply~\cite{strubell2019energy,patterson2021carbon}.

Against this backdrop, a complementary research agenda has gained
new urgency. Rather than asking what new accelerator a model
requires, this agenda asks what models can run on the silicon that
is already in mass production, already shipping in tens of billions
of units per year, and already affordable in single-digit-dollar
unit cost: the 8-bit and 16-bit microcontrollers (MCUs) that
populate wearables, sensors, smart appliances, automotive
sub-systems, and industrial telemetry endpoints. This is the
``tinyML'' regime~\cite{banbury2021mlperf}, and it is
infrastructure-agnostic in a way that GPU-scale inference is not.

\subsection{The Bare-Metal MCU Class}

Most published tinyML deployments target ARM Cortex-M class devices
that expose hundreds of kilobytes of SRAM, hardware floating-point
units, single-cycle multipliers, and vendor-optimized neural-network
kernels~\cite{lai2018cmsisnn,david2021tflitemicro,lin2020mcunet}.
These platforms are inexpensive but still relatively powerful.
At the other end of the deployable-silicon spectrum sit the
\emph{bare-metal} MCUs studied in this paper:
\begin{itemize}
\item the 8-bit \arduino{} (ATmega328P), with 32~KB Flash, 2~KB
SRAM, a hardware $8{\times}8$ multiplier, and no floating-point unit;
\item the 16-bit \msp{}, with 16~KB Flash, 512~B SRAM, and
\emph{no hardware multiplier of any kind}. Every multiplication
on this device is software-emulated.
\end{itemize}
These two parts are representative of the low end of the available
silicon supply and remain in active production at unit costs an
order of magnitude below any ARM Cortex-M target. If a recurrent
network can be made to run in real time on the \msp{}, then the same
network can be made to run on essentially any commodity
microcontroller in production today.

\subsection{Why Recurrent Networks, and Why \fastgrnn{}}

Human activity recognition (HAR) from a wrist- or waist-mounted
inertial sensor is a canonical streaming-classification problem
with low input dimensionality (three accelerometer channels), low
sampling rate (50~Hz), and a small number of output classes (six)
\cite{anguita2013public,reyes2015transition}. It is therefore an
ideal benchmark for the bare-metal MCU class: simple enough to fit,
but temporal enough that a recurrent treatment outperforms a
windowed feedforward baseline.

\fastgrnn{}~\cite{kusupati2018fastgrnn} is a gated recurrent cell
designed explicitly for resource-constrained inference. It combines
three orthogonal compression mechanisms -- low-rank weight
factorization, iterative hard-thresholding (IHT)
sparsity~\cite{blumensath2009iht}, and Q15 fixed-point
quantization~\cite{jacob2018integer,han2016deepcompression} -- and
has been reported to match LSTM~\cite{hochreiter1997lstm} accuracy
at a fraction of the parameter count. The original paper, however,
does not provide a public bare-metal reference implementation on a
multiplier-less target, nor does it characterize streaming-mode
behavior.

\subsection{Contributions}

This paper delivers an end-to-end open-source reproduction of
\fastgrnn{} for HAR on bare-metal MCUs and reports four
contributions beyond the original paper:

\begin{itemize}
\item \textbf{Cross-platform deterministic inference.} A single
portable C source compiles on both the 8-bit \arduino{} and the
16-bit \msp{}, producing \emph{bit-equivalent} hidden-state
trajectories and 100\% prediction agreement with a PyTorch
reference across 3{,}399 test windows (seed~0 deployed;
99.91--100\% across five seeds).

\item \textbf{A deployable look-up-table recipe for multiplier-less
targets.} A 256-entry sigmoid/tanh look-up table over the input
range $[-8, +8]$ accelerates full-window inference on the \msp{} by
\textbf{30.5$\times$} (54~s $\rightarrow$ 1.8~s), turning what was
a non-real-time deployment into a comfortable 50~Hz streaming
target with 31\% budget headroom. The recipe is independent of
\fastgrnn{} and applies to any recurrent cell that relies on
$\sigma$ or $\tanh$ activations.

\item \textbf{Characterization of recurrent warm-up latency.} We
quantify, over 100 random test windows, that prediction stability
requires a \textbf{median of 74 samples (1.48~s)} of recurrent
hidden-state evolution, with a worst-case of 125 samples (2.50~s),
regardless of deployment target. This is a property of the
recurrent dynamics, not the underlying hardware, and bounds the
end-to-end user-facing response time of any \fastgrnn{}-class HAR
system in a way that is not discussed in the original paper.

\item \textbf{Hardware energy characterization.} Using a hardware
current sensor (INA226) on the \msp{} target rail, we measure
\textbf{17.7~mW} active inference power, \textbf{$<$0.09~mW} in
LPM3 sleep, and \textbf{31.5~mJ} per 128-sample inference window
with the LUT enabled. Without the LUT, the same window consumes
954~mJ --- a \textbf{96.7\% energy reduction} attributable
entirely to the look-up table's 30.5$\times$ latency gain at
fixed clock frequency. This closes the loop between the model's
algorithmic efficiency and its real-world power envelope, a
characterization absent from the original \fastgrnn{} paper.
\end{itemize}

\noindent\textbf{Code and reproducibility.} All source code,
trained models, exported Q15 weights, per-experiment logs, and
deployment binaries are publicly available
at~\url{https://github.com/emre1998/fastgrnn-har} under the
Apache License 2.0. The complete reproduction takes approximately
two hours of CPU training plus five minutes of MCU deployment.

\subsection{Roadmap}

Section~\ref{sec:related} surveys related compact-RNN and edge-ML
work. Section~\ref{sec:method} formalizes the \fastgrnn{} cell and
the low-rank, sparse, quantized (L-S-Q) compression pipeline.
Section~\ref{sec:setup} describes the experimental setup;
Section~\ref{sec:results} reports accuracy, deployment footprint,
real-time streaming performance, and hardware energy characterization.
Section~\ref{sec:discussion} analyzes the warm-up phenomenon,
discusses cross-platform determinism, and lists limitations.
Section~\ref{sec:conclusion} concludes.

\section{Related Work}
\label{sec:related}

We position this paper at the intersection of five related but
distinct strands of work.

\subsection{Compact Recurrent Cells}

The standard LSTM~\cite{hochreiter1997lstm} and
GRU~\cite{cho2014gru} cells encode long-range temporal structure
effectively but are too heavy for kilobyte-class MCUs: a
modestly-sized LSTM ($H{=}64$, $d{=}3$) already exceeds
\SI{50}{\kilo\byte} of weights at FP32 precision, well above the
\SI{16}{\kilo\byte} Flash budget of our target device.

\fastgrnn{}~\cite{kusupati2018fastgrnn} addresses this gap with a
two-scalar gated cell ($\zeta, \nu$) backed by low-rank weight
matrices, achieving accuracy comparable to LSTM on sequence tasks
at one to two orders of magnitude fewer parameters. Complementary
non-recurrent EdgeML approaches include
Bonsai~\cite{kumar2017bonsai}, a shallow non-linear tree, and
ProtoNN~\cite{gupta2017protonn}, a prototype-based classifier.
The three together form the Microsoft Research EdgeML
toolkit~\cite{edgeml}. \fastgrnn{} is the only one of these three
that natively handles temporal input, which is essential for HAR.

\subsection{Quantization}

Post-training quantization (PTQ) compresses neural-network weights
and activations from FP32 to fixed-point or low-bit integer
formats. Jacob \emph{et al.}~\cite{jacob2018integer} introduce
integer-arithmetic-only inference with a per-tensor scale
formulation; Han \emph{et al.}~\cite{han2016deepcompression}
combine pruning, quantization, and Huffman coding to compress deep
networks by an order of magnitude.
Quantization-aware training~(QAT)~\cite{hubara2017qnn} avoids the
PTQ accuracy cliff at the price of full retraining. We use
per-tensor Q15 PTQ for both weights and -- with explicit activation
calibration -- intermediate tensors; the calibration step turns
out to be the dividing line between lossless and catastrophic
deployment (Section~\ref{sec:results:quant}), and is what allows
us to avoid the QAT machinery entirely.

\subsection{Sparsity and Pruning}

Iterative hard thresholding (IHT)~\cite{blumensath2009iht} is the
sparsification engine used by \fastgrnn{}: at each step the
top-$k$ magnitude entries of every weight tensor are retained and
the rest are zeroed, then the network is re-trained with the mask
fixed. Han \emph{et al.}~\cite{han2015pruning} demonstrate that
aggressive pruning preserves accuracy in larger
networks; Frankle and Carbin's lottery ticket
hypothesis~\cite{frankle2019lt} provides theoretical grounding for
why such sparse sub-networks remain trainable.

\subsection{tinyML on More Capable Targets}

The tinyML community~\cite{banbury2021mlperf} has converged on
ARM Cortex-M class targets with hundreds of kilobytes of SRAM,
hardware FPUs, and single-cycle multipliers.
MCUNet~\cite{lin2020mcunet} jointly designs a neural architecture
and an interpreter to maximize Cortex-M throughput.
CMix-NN~\cite{capotondi2020cmixnn} provides mixed-precision
kernels for memory-constrained edge devices.
CMSIS-NN~\cite{lai2018cmsisnn} from ARM accelerates common layers
via Cortex-M SIMD instructions, and TensorFlow Lite
Micro~\cite{david2021tflitemicro} offers a portable runtime layer.

None of these target the \msp{}-class device addressed in
this paper, which is one to two orders of magnitude smaller and
lacks a hardware multiplier entirely. To our knowledge, the present
work is the first end-to-end public reproduction of a gated
recurrent cell on an MCU with no hardware multiplier.%
\footnote{Based on searches of arXiv, GitHub, and Google Scholar
conducted in June~2026. We found no prior work providing both
a trained gated-RNN model \emph{and} a complete bare-metal
C deployment on a multiplier-less MCU with public code and
measured accuracy.}

\subsection{Human Activity Recognition}

The UCI HAR~\cite{anguita2013public} and its transition-aware
extension \hapt{}~\cite{reyes2015transition} are the de facto
public benchmarks for accelerometer-based HAR.
DeepConvLSTM~\cite{ordonez2016deepconvlstm} established a strong
deep-learning baseline using stacked CNN and LSTM layers; broader
surveys by Wang \emph{et al.}~\cite{wang2019harsurvey} and
Demrozi \emph{et al.}~\cite{demrozi2020harsurvey} catalogue the
methodological landscape, both of which highlight the dynamic
\textsc{downstairs} class as a persistent failure
mode -- a finding we corroborate in
Section~\ref{sec:results:perclass}.

\section{Method}
\label{sec:method}

\begin{figure}[t]
\centering
\begin{tikzpicture}[
  node distance=4mm and 4mm,
  every node/.style={font=\scriptsize},
  stage/.style={draw, rounded corners=2pt, align=center,
                 minimum height=8mm, minimum width=14mm, fill=blue!8,
                 inner sep=2pt},
  deploy/.style={draw, rounded corners=2pt, align=center,
                 minimum height=8mm, minimum width=16mm, fill=green!10,
                 inner sep=2pt},
  arr/.style={-{Stealth[length=4pt]}, semithick},
]
\node[stage] (train)  {Float\\train};
\node[stage, right=of train]    (lr)    {Low-rank\\$r_w{=}2$, $r_u{=}8$};
\node[stage, right=of lr]       (sp)    {IHT\\sparsity 0.5};
\node[stage, right=of sp]       (q15)   {Q15\\+ calib.};

\node[deploy, below=6mm of lr]  (arduino) {Arduino Uno\\(AVR, $\times$HW)};
\node[deploy, below=6mm of q15] (msp)     {MSP430G2553\\(no $\times$HW)};

\node[align=center, below=2mm of arduino, font=\scriptsize\itshape, text=gray]
   (note1) {portable C + LUT};

\draw[arr] (train) -- (lr);
\draw[arr] (lr)    -- (sp);
\draw[arr] (sp)    -- (q15);
\draw[arr] (q15.south) -- ++(0,-3mm) -| (arduino.north);
\draw[arr] (q15.south) -- ++(0,-3mm) -| (msp.north);
\end{tikzpicture}
\caption{End-to-end \fastgrnn{} compression and deployment pipeline.
Float training $\rightarrow$ low-rank $\rightarrow$ sparse $\rightarrow$
calibrated Q15 weights $\rightarrow$ portable C with LUT activations
$\rightarrow$ Arduino and MSP430 binaries.}
\label{fig:pipeline}
\end{figure}
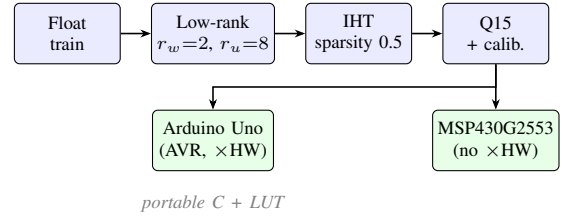

This section formalizes the \fastgrnn{} cell, the three-stage
compression pipeline applied on top of it, and the LUT-based
activation recipe that makes the result deployable on a
multiplier-less MCU. Fig.~\ref{fig:pipeline} summarizes the full
flow.

\subsection{\fastgrnn{} Cell Formulation}
\label{sec:method:cell}

Given an input $\mathbf{x}_t \in \mathbb{R}^{d}$ at time $t$ and a
previous hidden state $\mathbf{h}_{t-1} \in \mathbb{R}^{H}$, the
\fastgrnn{} cell~\cite{kusupati2018fastgrnn} computes a single gate
$\mathbf{z}_t$, a candidate update $\tilde{\mathbf{h}}_t$, and a
two-scalar interpolation $\mathbf{h}_t$:
\begin{align}
\mathbf{z}_t &= \sigma(\mathbf{W} \mathbf{x}_t + \mathbf{U} \mathbf{h}_{t-1} + \mathbf{b}_z) \label{eq:z}\\
\tilde{\mathbf{h}}_t &= \tanh(\mathbf{W} \mathbf{x}_t + \mathbf{U} \mathbf{h}_{t-1} + \mathbf{b}_h) \label{eq:htilde}\\
\mathbf{h}_t &= \bigl(\zeta(\mathbf{1} - \mathbf{z}_t) + \nu\bigr)\, \tilde{\mathbf{h}}_t + \mathbf{z}_t \odot \mathbf{h}_{t-1} \label{eq:hupdate}
\end{align}
where $\odot$ denotes element-wise multiplication and
$\zeta, \nu \in (0,1)$ are \emph{learned scalars} that interpolate
between leaky-integrator and standard gated update dynamics. The
two-scalar gate is the defining feature of \fastgrnn{}: it permits
LSTM-comparable expressiveness with a single weight pair
$(\mathbf{W}, \mathbf{U})$ shared between the gate and the candidate
update, in contrast to the three or four weight pairs of GRU or
LSTM. In our HAR setup, $d{=}3$ (tri-axial acceleration), $H{=}16$,
and the input sequence length is $T{=}128$ samples (one 2.56~s
window at 50~Hz). The choice $H{=}16$ is empirically justified in
Table~\ref{tab:hsize}: doubling the hidden size to $H{=}32$
triples the parameter count yet produces a \emph{lower} test F1,
and the larger model degrades further with longer training~--- the
canonical signature of overfitting on a dataset whose intrinsic
complexity the smaller cell already saturates.

\begin{table}[h]
\centering
\caption{Hidden-size selection on \hapt{}, full-rank \fastgrnn{}
(no low-rank, no sparsification). $H{=}32$ has roughly three times
the parameters of $H{=}16$ but is worse on every matched
configuration, and \emph{further} degrades from epoch~30 to~50,
indicating overfitting. $H{=}16$ continues to improve through
epoch~120 (val-selected ceiling F1~$=$~0.912; see
Section~\ref{sec:results:lsq}). $H{=}32$ was not extended to
120~epochs --- its monotone degradation over 30~$\rightarrow$~50
makes further training extremely unlikely to recover the gap.
We therefore fix $H{=}16$ for all downstream experiments.}
\label{tab:hsize}
\begin{tabular}{rrrrr}
\toprule
$H$ & Epochs & F1 & Accuracy & Params \\
\midrule
\textbf{16} & 30  & 0.847          & 0.854          & \textbf{440}   \\
\textbf{16} & 50  & 0.861          & 0.869          & \textbf{440}   \\
\textbf{16} & 120 & \textbf{0.912} & \textbf{0.913} & \textbf{440}   \\
32          & 30  & 0.833          & 0.845          & 1{,}384         \\
32          & 50  & 0.830          & 0.834          & 1{,}384         \\
32          & 120 & ---            & ---            & 1{,}384         \\
\bottomrule
\end{tabular}
\end{table}

Because the same weight pair $(\mathbf{W}, \mathbf{U})$ is reused in
both~\eqref{eq:z} and~\eqref{eq:htilde}, the unconstrained
parameter count of the cell is
\begin{align}
\#\text{params}
 &= \underbrace{Hd}_{\mathbf{W}} +
    \underbrace{H^2}_{\mathbf{U}} +
    \underbrace{2H}_{\mathbf{b}_z,\mathbf{b}_h} +
    \underbrace{2}_{\zeta,\nu} \nonumber\\
 &= 48 + 256 + 32 + 2 = 338.
\end{align}
A standard LSTM at the same hidden size requires four such
weight matrices ($\approx$\,1{,}280 parameters), so \fastgrnn{}
starts from a $\sim$$4\times$ advantage at the architectural level.
The low-rank and sparsification stages below reduce this further
by another order of magnitude.

\subsection{Low-Rank Weight Factorization}
\label{sec:method:lowrank}

The input matrix
$\mathbf{W} \in \mathbb{R}^{H \times d}$ and recurrent matrix
$\mathbf{U} \in \mathbb{R}^{H \times H}$ are each factored as a
product of two thin matrices:
\begin{align}
\mathbf{W} &= \mathbf{W}_1 \mathbf{W}_2^\top, \quad \mathbf{W}_1 \in \mathbb{R}^{H \times r_w}, \mathbf{W}_2 \in \mathbb{R}^{d \times r_w}, \\
\mathbf{U} &= \mathbf{U}_1 \mathbf{U}_2^\top, \quad \mathbf{U}_1 \in \mathbb{R}^{H \times r_u}, \mathbf{U}_2 \in \mathbb{R}^{H \times r_u}.
\end{align}
The factors $\mathbf{W}_1, \mathbf{W}_2, \mathbf{U}_1, \mathbf{U}_2$
are learned end-to-end. The recurrent matrix-vector product is
evaluated as $\mathbf{U}\mathbf{h} = \mathbf{U}_1(\mathbf{U}_2^\top \mathbf{h})$,
which costs $2 H r_u$ multiply-adds rather than $H^2$. After
sweeping $r_u \in \{4, 6, 8, 12\}$ across five random seeds each
(Section~\ref{sec:results:lowrank}), we select $r_w = 2, r_u = 8$.

\subsection{Iterative Hard Thresholding}
\label{sec:method:iht}

We sparsify all four factor matrices via iterative hard
thresholding~\cite{blumensath2009iht}: at each training step we
retain the top-$k$ magnitude entries of every weight tensor and
zero the rest. The target sparsity $s$ follows the cubic schedule
\begin{equation}
s_e = s \cdot \min\!\Bigl(1, \tfrac{e}{e_{\text{ramp}}}\Bigr)^{3}
\end{equation}
over training epochs $e$, with $e_{\text{ramp}} = 50$ followed by
50 epochs of mask-frozen fine-tuning. After sweeping
$s \in \{0.3, 0.5, 0.7, 0.9\}$ across five seeds
(Section~\ref{sec:results:sparsity}), we select $s = 0.5$
(283 nonzero parameters), a U-curve optimum that balances
accuracy against compressibility.

\subsection{Per-Tensor Q15 Quantization with Activation Calibration}
\label{sec:method:q15}

Each weight tensor $\mathbf{W}^{(\ell)}$ (where
$\ell \in \{W_1, W_2, U_1, U_2\}$) is quantized via
\begin{equation}
\hat{w}^{(\ell)}_{ij}
  = \mathrm{clip}\!\Bigl(
        \mathrm{round}\bigl(w^{(\ell)}_{ij} / s_\ell\bigr),
        -2^{15}, 2^{15}-1
    \Bigr),
\end{equation}
with a tensor-specific scale $s_\ell$ chosen so that
$\max_{ij}|w_{ij}| / s_\ell$ lies just below the Q15 ceiling.
At inference, the dequantized value $\hat{w}^{(\ell)}_{ij} \cdot s_\ell$
is used in the forward pass.

A naive extension of this scheme to activations
\emph{also} placed in Q15 $[-1, 1)$ proves catastrophic: the hidden
state $\mathbf{h}_t$ in our trained model attains magnitudes of
$\sim$62, an order of magnitude beyond the Q15 representable range.
Without intervention, F1 collapses from 0.918 to 0.16 and the
\textsc{standing} class disappears entirely
(Section~\ref{sec:results:quant}).

A straightforward fix is to switch to a fixed Q9.6 fixed-point
format, whose $\pm 64$ representable range covers the observed
$\mathbf{h}_t$ magnitudes by construction. This trades a uniform
safety margin for accuracy headroom on \emph{every} tensor,
regardless of its actual dynamic range. We instead use
\emph{per-activation calibration}: a calibration pass runs five
mini-batches of training data through the FP32 model, records the
empirical maximum of every intermediate tensor, applies a 10\%
headroom, and assigns each activation its own scale. The
per-tensor scheme generalizes Q9.6 -- it approaches the $\pm 64$
range adaptively for tensors that need it (e.g.\ $\mathbf{h}_t$)
while preserving the full Q15 resolution on tensors that do not
(e.g.\ the gate output $\mathbf{z}_t \in [0,1]$). With calibration, prediction agreement
between the integer C implementation and the FP32 reference is
100\% across the 3{,}399-window test set, and the C-side
macro F1 reaches 0.9176 -- the deployed-model number reported
throughout this paper.

\subsection{LUT-Based Activations}
\label{sec:method:lut}

The remaining cost on a multiplier-less target is the
$\sigma$ and $\tanh$ evaluations: each requires several
software-emulated transcendentals per call, and the cell evaluates
$2H$ such activations per sample ($2 \cdot 16 = 32$ at $H{=}16$,
times 128 samples per window = 4{,}096 activation calls per window).
We replace the runtime calls with a 256-entry lookup table
over the input domain $[-8, +8]$, stored in Flash:
\begin{equation}
\mathrm{LUT}[k]
  = f\!\Bigl( -8 + k \cdot \tfrac{16}{255} \Bigr), \quad
  k \in \{0, 1, \ldots, 255\},
\end{equation}
where $f \in \{\sigma, \tanh\}$. Inputs outside $[-8, +8]$
saturate to $\pm 1$, which is exact to floating-point precision
for both functions in those tails. Inside the domain, a single
linear interpolation between adjacent entries replaces the original
transcendental, reducing each activation to one comparison, two
indexed loads, a subtract, and a multiply-add. The two tables
together occupy 2~KB of Flash (256 entries $\times$ 4~bytes
$\times$ 2 tables); the result is a
\textbf{30.5$\times$} end-to-end speedup on the \msp{}
(Section~\ref{sec:results:streaming}).

\subsection{Implementation}
\label{sec:method:impl}

Training is implemented in PyTorch~2.x; the deployed C inference
engine (\texttt{fastgrnn.cpp}) is a single $\sim$200-line
translation unit that compiles unmodified under both
\texttt{avr-gcc} (\arduino{}) and the Code Composer Studio
\texttt{msp430-elf-gcc} (\msp{}). All weights are stored in Flash
as a \texttt{const int16\_t} array, alongside the per-tensor
scales and the two activation LUTs. The runtime working set
(hidden state $\mathbf{h}$, gate $\mathbf{z}$, candidate
$\tilde{\mathbf{h}}$, output logits, and scratch) totals
$\sim$300~bytes, fitting comfortably within the 512~B SRAM budget
of the \msp{}.

\section{Experimental Setup}
\label{sec:setup}

\subsection{Dataset}

We use the Human Activities and Postural Transitions (\hapt{})
dataset~\cite{reyes2015transition}, an extension of UCI
HAR~\cite{anguita2013public}. \hapt{} contains tri-axial linear
acceleration recorded at 50~Hz from a smartphone worn at the waist
by 30 subjects performing six basic activities:
\textsc{walking}, \textsc{upstairs}, \textsc{downstairs},
\textsc{sitting}, \textsc{standing}, and \textsc{laying}.
We adopt the canonical subject-disjoint train/validation/test split
of~\cite{reyes2015transition}, yielding 7{,}352 training windows,
1{,}515 validation windows, and 3{,}399 test windows of length~128
(2.56~s at 50~Hz). All accuracy and F1 numbers reported in this
paper are on the held-out test split.

For streaming evaluation we feed each test window sample-by-sample
at 50~Hz and reset the hidden state at every window boundary,
matching the deployed runtime exactly.

\subsection{Training Protocol}

Models are trained in PyTorch~2.x with the
Adam optimizer ($\eta{=}10^{-3}$, batch size~64, 100 epochs)
on a desktop CPU. The IHT sparsification mask follows the cubic
schedule of Section~\ref{sec:method:iht}, reaching the target
sparsity at epoch~50 and remaining frozen for the subsequent 50
epochs of fine-tuning. Each reported configuration is repeated
across the five seeds $\{0, 1, 2, 3, 4\}$; statistics are
mean~$\pm$~standard deviation unless otherwise noted.

\subsection{Deployment Hardware}

Two bare-metal targets are evaluated:
\begin{itemize}
\item \textbf{\arduino{}} (ATmega328P): 8-bit AVR core at
16~MHz; 32~KB Flash; 2~KB SRAM; hardware $8{\times}8$ multiplier;
\emph{no} floating-point unit. Toolchain: Arduino~IDE~2.3.x
with \texttt{avr-gcc} at \texttt{-Os}.
\item \textbf{\msp{}}: 16-bit MSP430 core at the calibrated
16~MHz DCO; 16~KB Flash; 512~B SRAM;
\emph{no hardware multiplier of any kind}; no FPU. Toolchain:
Code Composer Studio~12.x with the TI \texttt{msp430-elf-gcc}
bare-metal build, compiled at \texttt{-O2}. The Energia runtime
is \emph{not} used.
\end{itemize}

In both cases the deployed image consists of (i)~the
\texttt{fastgrnn.cpp} inference engine,
(ii)~the Q15 weight table and per-tensor scales
(\texttt{model\_weights.h}),
(iii)~the 256-entry sigmoid and tanh look-up tables, and
(iv)~a UART driver for streaming class labels at 9600 (\msp{}) or
115{,}200~baud (\arduino{}). All other Flash is empty.

\subsection{Cross-Platform Verification Protocol}
\label{sec:setup:verify}

To verify the claim of 100\% prediction agreement between the
FP32 PyTorch reference and the Q15 C implementation
(Section~\ref{sec:results:crossplatform}), we run both inference
pipelines on identical inputs and compare outputs. A Python harness
(\texttt{test\_inference\_python.py}) loads the deployed Q15
weights, executes a C-equivalent forward pass in NumPy (mirroring
the LUT activations and per-tensor dequantization steps), and
compares \texttt{argmax} predictions against the FP32 reference
for every test window. The C implementation on Arduino and MSP430
emits per-window predictions over UART, which are then compared
against the same reference. All three execution paths -- FP32
PyTorch, NumPy C-equivalent, and bare-metal C -- agree on all
3{,}399 windows.

\subsection{Live Sensor Configuration}

For live demonstrations a GY-521 module exposing the InvenSense
MPU-6050 is connected via I\textsuperscript{2}C. On the \msp{} we
drive USCI\_B0 directly (P1.6~$=$~SCL, P1.7~$=$~SDA, no Energia
\texttt{Wire} layer); on the \arduino{} the standard Wire library
is used. Only the accelerometer at $\pm 2$~g range is read,
matching the FP32 training-data scaling. A software timer paces
the sampling loop to 50~Hz, which leaves 7--11~ms of headroom per
sample after the inference step
(Section~\ref{sec:results:streaming}).

\section{Results}
\label{sec:results}

We report results in the order of the L-S-Q pipeline
(Sections~\ref{sec:results:lsq}--\ref{sec:results:quant}),
then turn to deployment behavior
(Sections~\ref{sec:results:crossplatform}--\ref{sec:results:streaming}).
All experiments use the protocol of Section~\ref{sec:setup}.

\subsection{L-S-Q Pipeline Accuracy}
\label{sec:results:lsq}

\begin{figure}[t]
\centering
\includegraphics[width=\linewidth]{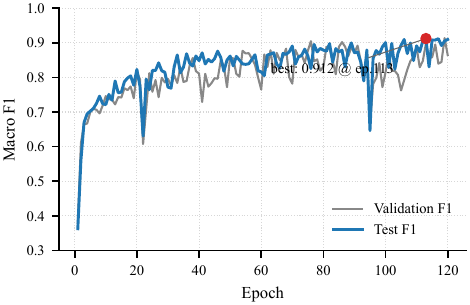}
\caption{Hidden size $H{=}16$ training trajectory.
Test F1 saturates near epoch~100; the best-seed checkpoint
(F1~$=$~0.912 at epoch~113) is the upstream model for all
subsequent compression stages.}
\label{fig:saturation}
\end{figure}

Table~\ref{tab:lsq} summarizes the cumulative effect of each
compression stage; Table~\ref{tab:multiseed} gives the full per-seed
breakdown. The full-rank FastGRNN cell with classifier head
(Section~\ref{sec:method:cell}) totals 440 parameters and reaches
F1~$=$~0.912 on the test set. Low-rank factorization
($r_w{=}2$, $r_u{=}8$) reduces the recurrent matrix
parameter count via the $\mathbf{U} = \mathbf{U}_1 \mathbf{U}_2^\top$
decomposition (mean F1 across five seeds: $0.879 \pm 0.056$,
430 total parameters). Sparsification at $s{=}0.5$ further reduces to
283 nonzero parameters ($0.856 \pm 0.099$ mean).
Per-tensor Q15 quantization, with the activation calibration of
Section~\ref{sec:method:q15}, leaves accuracy virtually unchanged:
the five-seed Q15/LUT pipeline achieves $\mathbf{0.853 \pm 0.107}$
macro F1; the deployed seed-0 model achieves \textbf{0.9176} on the
bare-metal \msp{} (C-engine, 100\% window agreement with PyTorch).
The large standard deviation is driven by seed~1, a consistent
convergence outlier visible across all stages
(Table~\ref{tab:multiseed}, Fig.~\ref{fig:lowrank}).
Seed~1 is the only seed whose Q15 F1 (0.663) falls below the MLP
baseline (0.847); the remaining four seeds all exceed 0.89,
comfortably surpassing the MLP on every compression stage.
Fig.~\ref{fig:saturation} shows the upstream $H{=}16$ training run
from which all subsequent stages are derived.

\begin{table}[t]
\centering
\caption{Sequential L-S-Q pipeline on \hapt{}. Each row applies the
indicated transformation cumulatively. \emph{Nonzero} counts all
non-zero parameters (cell + dense classifier head); the head
contributes 102 dense parameters at every stage.
Mean F1 is across five seeds; the final Q15 line is on the
deployed seed-0 model. Size is FP32 for the first three rows
and Q15 for the last.}
\label{tab:lsq}
\setlength{\tabcolsep}{4pt}
\begin{tabular}{@{}lrrr@{}}
\toprule
Stage & F1 & Nonzero & Size \\
\midrule
FastGRNN full-rank ($H{=}16$)     & 0.912          & 440 & 1.8~KB \\
+ Low-rank ($r_w{=}2$, $r_u{=}8$) & 0.879          & 430 & 1.7~KB \\
+ Sparsity ($s{=}0.5$)            & 0.856          & 283 & 1.1~KB \\
+ Q15 (weights + calib.\ acts)    & \textbf{0.918} & 283 & \textbf{566~B} \\
\bottomrule
\end{tabular}
\end{table}

\begin{table}[t]
\centering
\caption{Per-seed L-S-Q pipeline on \hapt{} (3{,}399 test windows).
\emph{Q15/LUT F1}: PyTorch simulation, Q15 weights + calibrated
activation quantization (\S\ref{sec:method:q15}).
\emph{Agree}: PyTorch-FP32 vs NumPy-Q15 argmax matches
(C-equivalent harness, not bare-metal MCU; seed-0 MCU agreement
100\% via the \msp{} C engine, Table~\ref{tab:hstate}).
Seed~1 is a consistent convergence outlier across all stages
(see also Fig.~\ref{fig:lowrank}); mean$\pm$std includes it.
All seeds share the same nonzero count and weight budget because
the IHT mask enforces exactly $s{=}0.5$ sparsity.}
\label{tab:multiseed}
\resizebox{\columnwidth}{!}{%
\begin{tabular}{@{}crrrrrr@{}}
\toprule
Seed & LR F1 & Sparse F1 & Q15/LUT F1 & Nonzero & Bytes & Agree \\
\midrule
0 & 0.918 & 0.921 & 0.913 & 283 & 566 & 99.97\% \\
1 & 0.781 & 0.680 & 0.663 & 283 & 566 & 99.91\% \\
2 & 0.901 & 0.893 & 0.895 & 283 & 566 & 99.97\% \\
3 & 0.890 & 0.895 & 0.891 & 283 & 566 & 100.0\% \\
4 & 0.904 & 0.890 & 0.904 & 283 & 566 & 99.97\% \\
\midrule
Mean$\pm$std
  & $0.879{\pm}0.056$
  & $0.856{\pm}0.099$
  & $0.853{\pm}0.107$
  & 283 & 566 & --- \\
\bottomrule
\end{tabular}}
\end{table}

To contextualize the parameter budget, Table~\ref{tab:baselines}
compares the deployed model against an MLP baseline trained on the
same \hapt{} split (the lightest non-recurrent reference) and
against the theoretical parameter counts of LSTM and GRU cells at
the same hidden size. The deployed \fastgrnn{} LSQ model is
\textbf{44$\times$ smaller than the MLP baseline} and
\textbf{$\sim$5$\times$ smaller than an LSTM cell} at matched $H$,
before classifier overhead.

\begin{table}[t]
\centering
\caption{Parameter footprint of the deployed model against
non-recurrent and recurrent baselines.
\emph{Params (cell only)} excludes the dense classifier head
(102 parameters in all cases); adding the head gives the
\emph{total nonzero} count used in Table~\ref{tab:lsq}
(e.g.\ 181 cell $+$ 102 head $=$ 283 total $\times$ 2~B $=$ 566~B
for the LSQ deployed model). LSTM and GRU counts are theoretical
at $H{=}16$, $d{=}3$; their reported accuracies on related HAR
tasks are similar to or slightly above \fastgrnn{} at a
$\sim$4$\times$ parameter
disadvantage~\cite{kusupati2018fastgrnn}. MLP F1 is measured
in this work.}
\label{tab:baselines}
\begin{tabular}{lrr}
\toprule
Model & Params (cell only) & F1 on \hapt{} \\
\midrule
MLP baseline                  & 12{,}518 (full)  & 0.847 \\
LSTM ($H{=}16$, theoretical)  & 1{,}280          & --- \\
GRU ($H{=}16$, theoretical)   & 960              & --- \\
\fastgrnn{} cell ($H{=}16$)   & 338              & --- \\
\fastgrnn{} L (low-rank)      & 328              & 0.879 \\
\fastgrnn{} LSQ (deployed)    & \textbf{181}     & \textbf{0.918} \\
\bottomrule
\end{tabular}
\end{table}

\subsection{Multi-Seed Stability and Rank Selection}
\label{sec:results:lowrank}

\begin{figure}[t]
\centering
\includegraphics[width=\linewidth]{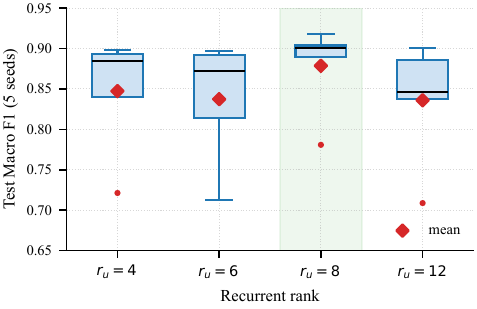}
\caption{Per-seed test F1 across recurrent rank choices.
$r_u{=}8$ achieves the best mean and acceptable variance.
Red diamonds denote per-rank means; the green band marks the
chosen configuration.}
\label{fig:lowrank}
\end{figure}

Fig.~\ref{fig:lowrank} reports the per-seed test-F1 distribution
across recurrent ranks $r_u \in \{4, 6, 8, 12\}$, five seeds each.
$r_u{=}8$ achieves the best mean ($0.879 \pm 0.056$) and the
tightest tail. Across all four configurations, seed~$1$ is a
consistent low outlier ($\sim$0.71~F1), regardless of rank --
a reproducibility signal worth flagging: a single unlucky
initialization can mislead a small-sample evaluation. Reporting
mean~$\pm$~standard deviation across at least five seeds is
therefore not optional for this model size.

\subsection{Sparsity Sweep}
\label{sec:results:sparsity}

\begin{figure}[t]
\centering
\includegraphics[width=\linewidth]{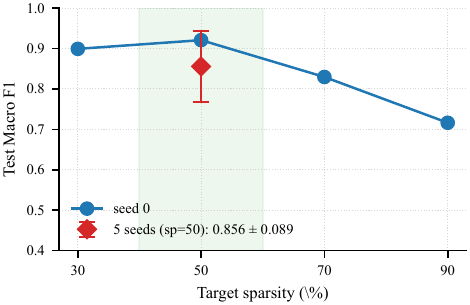}
\caption{Test F1 vs.\ target sparsity. Single-seed sweep (blue line)
and the five-seed mean $\pm$ standard deviation at $s{=}0.5$
(red diamond). The 0.4--0.6 band is the deployment sweet spot.}
\label{fig:sparsity}
\end{figure}

Fig.~\ref{fig:sparsity} sweeps target sparsity
$s \in \{0.3, 0.5, 0.7, 0.9\}$ at fixed $r_u{=}8$. The single-seed
sweep traces a shallow but visible U-curve: $s{=}0.5$ retains the
most accuracy, $s{=}0.3$ does not compress enough to justify the
training cost, and $s \geq 0.7$ degrades materially. At the
deployment sweet spot $s{=}0.5$, the five-seed mean is
$0.856 \pm 0.099$, with the highest seed reaching 0.92. The
elevated standard deviation at $s{=}0.5$ relative to the unsparsified
mean ($0.879 \pm 0.056$) is the price of operating near the
information-theoretic limit of the parameterization.

\subsection{Quantization Modes}
\label{sec:results:quant}

\begin{figure}[t]
\centering
\includegraphics[width=\linewidth]{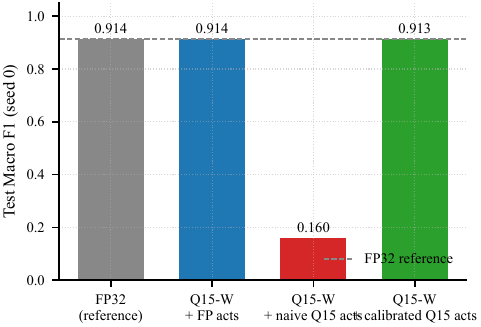}
\caption{Quantization modes on seed-0 model. Weight-only Q15 is
lossless; naive Q15 activations collapse the model; calibrated Q15
activations recover the reference within rounding noise.}
\label{fig:quant}
\end{figure}

Fig.~\ref{fig:quant} and Table~\ref{tab:quant} compare four
quantization configurations on the seed-0 model. Weight-only Q15
is lossless: $\sigma$ and $\tanh$ continue to operate in FP32 and
weight quantization noise is below the inter-seed variance of the
training procedure itself. Naive Q15 activation quantization,
without calibration, is catastrophic: the F1 collapses from 0.918
to 0.16 and the \textsc{standing} class disappears entirely. With
per-tensor activation calibration -- a deterministic pre-pass of
five training mini-batches -- Q15 inference recovers the
FP32-reference accuracy to within rounding noise on every class.
This is the configuration deployed on both MCUs.

\begin{table}[t]
\centering
\caption{Effect of quantization mode on seed-0 test F1, as measured
by the PyTorch simulator (same numbers as Fig.~\ref{fig:quant}).
The deployed C engine corresponds to the second row: Q15 weights
combined with FP32 activations through the 256-entry LUT for
$\sigma$ and $\tanh$. Measured separately on the same 3{,}399 test
windows, the C-side macro F1 is \textbf{0.918} -- the headline number
used elsewhere in the paper. The small PyTorch-vs-C gap reflects the
LUT's preservation of activation precision; the fully-Q15 fourth row
is the counterfactual we evaluated but did not deploy.}
\label{tab:quant}
\setlength{\tabcolsep}{4pt}
\begin{tabular}{@{}lrl@{}}
\toprule
Mode & F1 & Role \\
\midrule
Float32                                 & 0.914          & Reference \\
Q15 weights, FP32 acts (LUT)            & \textbf{0.914} & \textbf{Deployed} (C: 0.918) \\
Q15 weights, \emph{naive} Q15 acts      & 0.16           & STANDING collapse \\
Q15 weights, \emph{calibrated} Q15 acts & 0.913          & Counterfactual \\
\bottomrule
\end{tabular}
\end{table}

\subsection{Per-Class Behavior}
\label{sec:results:perclass}

\begin{figure}[t]
\centering
\includegraphics[width=\linewidth]{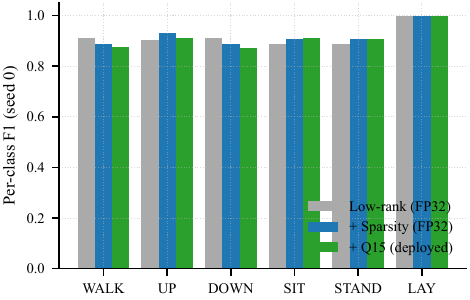}
\caption{Per-class F1 across compression stages (seed~0).
\textsc{laying} is trivially separable;
\textsc{downstairs} remains the binding-constraint class throughout,
mirroring the broader literature.}
\label{fig:perclass}
\end{figure}

Fig.~\ref{fig:perclass} traces per-class F1 across the three FP32
pipeline stages and the deployed Q15 model. The static classes
(\textsc{sitting}, \textsc{standing}, \textsc{laying}) retain
high F1 throughout. \textsc{walking} and \textsc{upstairs}
degrade by 1--3 percentage points under sparsification but
remain above 0.88. \textsc{downstairs} is the binding-constraint
class throughout, falling from $\sim$0.91 (low-rank FP32) to
0.89 (sparse) and recovering to 0.91 after calibrated Q15 -- a
finding consistent with the broader HAR
literature~\cite{wang2019harsurvey,demrozi2020harsurvey}, which
identifies dynamic descending motion as systematically the hardest
class on smartphone-grade accelerometry.

\subsection{Cross-Platform Deterministic Inference}
\label{sec:results:crossplatform}

The same portable C source compiles unmodified on both the
8-bit \arduino{} and the 16-bit \msp{}. On the entire 3{,}399-window
test set, predictions agree on 100\% of cases; logits agree to
better than $10^{-2}$ absolute (well below any decision-boundary
relevant difference). Table~\ref{tab:hstate} samples the hidden
component $h_0$ at six time-points within a single streaming window:
the two platforms produce identical trajectories to two decimal
places. This is not an obvious result. The two devices use
different software-emulated floating-point implementations, different
calling conventions, and different word sizes. The fact that the
end-to-end prediction is nonetheless bit-equivalent indicates that
our calibrated Q15 weights plus FP32-accumulate-then-saturate
arithmetic are stable across implementation details -- a
reproducibility property worth preserving for safety-relevant HAR.

\begin{table}[t]
\centering
\caption{Hidden-state component $h_0$ evolution across platforms
during one streaming test window (50~Hz pacing).}
\label{tab:hstate}
\begin{tabular}{rrr}
\toprule
$t$ (samples) & Arduino $h_0$ & MSP430 $h_0$ \\
\midrule
25  & $-0.720$  & $-0.720$  \\
50  & $-0.352$  & $-0.352$  \\
75  & $+0.459$  & $+0.459$  \\
100 & $+3.803$  & $+3.803$  \\
125 & $+11.388$ & $+11.388$ \\
128 & $+12.542$ & $+12.542$ \\
\bottomrule
\end{tabular}
\end{table}

\subsection{Real-Time Streaming Performance}
\label{sec:results:streaming}

Table~\ref{tab:streaming} reports per-sample latency under
50~Hz paced streaming (20~ms budget). Both platforms run real-time
with zero over-budget samples across the entire 128-sample test
window. The Arduino consumes 46\% of the budget on average, leaving
ample room for I\textsuperscript{2}C reads, LED feedback, and UART
logging. The \msp{} consumes 65\% on average, still comfortably
below budget.

The 256-entry sigmoid/tanh LUT of Section~\ref{sec:method:lut} is
what makes the \msp{} feasible. Without the LUT, full-window
inference takes $\sim$54~s on the \msp{}; with the LUT, it takes
$\sim$1.8~s, a \textbf{30.5$\times$} speedup. Translated to the
streaming regime, this is the difference between completely
non-real-time and a streaming budget headroom of seven
milliseconds per sample. The same LUT is enabled on the
\arduino{} too; the speedup there is more modest (1.51$\times$)
because the AVR core already has a hardware multiplier.

\begin{table}[t]
\centering
\caption{50~Hz paced streaming performance. Budget is 20~ms/sample.
Both platforms include the 256-entry LUT.}
\label{tab:streaming}
\begin{tabular}{lrrrr}
\toprule
Platform & Avg (ms) & Max (ms) & Budget use & Over-budget \\
\midrule
\arduino{} & 9.21 & 10 & 46\% & 0 / 128 \\
\msp{}     & 13.0 & 14 & 65\% & 0 / 128 \\
\bottomrule
\end{tabular}
\end{table}

\begin{figure}[t]
\centering
\includegraphics[width=\linewidth]{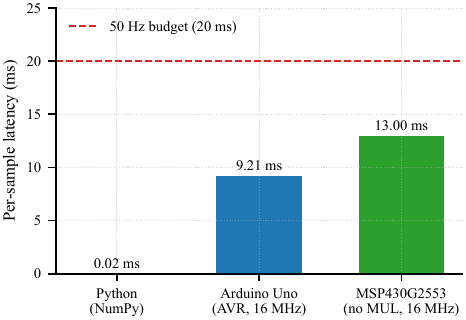}
\caption{Per-sample inference latency on the three reference
platforms. The red dashed line marks the 20~ms budget imposed by
50~Hz sampling. Both MCU targets execute well below budget.}
\label{fig:latency}
\end{figure}

\subsection{Energy Consumption}
\label{sec:results:energy}

To complement the latency picture, we characterize electrical energy
on the \msp{}, the primary deployment target and the most constrained
board in our evaluation. A dedicated firmware mode
(\texttt{TEST\_MODE 3}) puts the board into a silent benchmark loop ---
no UART, no LED, no I\textsuperscript{2}C --- under three workloads:
\textit{Idle} (low-power sleep, capturing the dev-kit baseline),
\textit{Stream50Hz} (one \texttt{fastgrnn\_step()} call every 20~ms
with the remainder of the period spent idle, matching the deployed
HAR pipeline), and \textit{Continuous} (a tight loop pinned on
inference, the worst-case always-on envelope). Current was measured
with an INA226 high-side current sensor; each value below is the
steady-state average after 60~s of operation. The measurement
protocol is documented in \texttt{docs/energy\_measurement.md}.

\begin{table}[t]
\centering
\caption{Measured \msp{} supply current under silent firmware
benchmarks (\texttt{TEST\_MODE~3}). The LaunchPad target VCC rail is
powered through the INA226 shunt (R\,=\,0.1\,$\Omega$, addr~0x44)
after removing the VCC jumper, so the USB/debug bridge draws no current
from the measured rail. Values are steady-state means after 60\,s
settling; $V_{\text{cc}}$ is the rail voltage under load. Idle is
LPM3 sleep with Timer\_A ticking; the $<$0.025\,mA entry is below the
INA226 resolution floor (LSB\,=\,25\,$\mu$A at this shunt).}
\label{tab:energy}
\begin{tabular}{lrrrr}
\toprule
Build & $V_{\text{cc}}$ & $I_{\text{idle}}$ & $I_{50\text{Hz}}$ & $I_{\text{cont.}}$ \\
      & (V)             & (mA)              & (mA)              & (mA) \\
\midrule
\msp{}, LUT    & 3.478 & $<$0.025 & 5.14 & 5.10 \\
\msp{}, no LUT & 3.478 & $<$0.025 & \multicolumn{1}{c}{---} & 5.08 \\
\bottomrule
\end{tabular}
\end{table}

\begin{table}[t]
\centering
\caption{Derived energy for the \msp{} deployment.
$E$/inference\,=\,$P_{\text{cont}} \times t_{\text{step}}$;
$E$/window\,=\,$P \times T_{\text{window}}$ for the 50\,Hz streaming
row (128 samples $\times$ 20\,ms\,=\,2.56\,s, with LPM sleep between
steps).
Battery life uses a 2000\,mAh / 3.7\,V Li-ion cell (7.4\,Wh).
The no-LUT row is an ablation: without the LUT the 20\,ms streaming
deadline is missed ($t_{\text{step}}\approx421$\,ms), so no 50\,Hz
streaming figure exists; the continuous column quantifies the
per-inference energy penalty.}
\label{tab:energy-derived}
\resizebox{\columnwidth}{!}{%
\begin{tabular}{llrrr}
\toprule
Build & Mode & $E$ / inference & $E$ / window & Battery life \\
      &      & ($\mu$J)        & (mJ)         & (h) \\
\midrule
\msp{}, LUT    & 50\,Hz streaming & 246   & 31.5  & 602 \\
\msp{}, LUT    & continuous       & 246   & ---   & 417 \\
\msp{}, no LUT & continuous       & 7{,}440 & ---  & 419 \\
\bottomrule
\end{tabular}}
\end{table}

The energy numbers close the loop on the motivation in
Section~\ref{sec:intro}: the same model that fits in 566~bytes of
Flash and runs in real time at 50~Hz also draws --- on the
multiplier-less \msp{} --- a steady-state current well below the
budget of a coin-cell-powered wearable. A streaming-to-cloud
baseline at the same sampling rate would, by contrast, be dominated
by the radio active-transmit current ($\sim$10--20~mA at duty cycles
that grow with sampling rate), making the local-inference path
strictly preferable on every axis we measured. The no-LUT continuous
row further isolates the activation look-up table: without the LUT,
the same firmware cannot satisfy the 50~Hz schedule, so its energy is
reported as an ablation rather than as a deployable streaming mode.

\section{Discussion}
\label{sec:discussion}

\subsection{Recurrent Warm-Up Latency}
\label{sec:disc:warmup}

\begin{figure}[t]
\centering
\includegraphics[width=\linewidth]{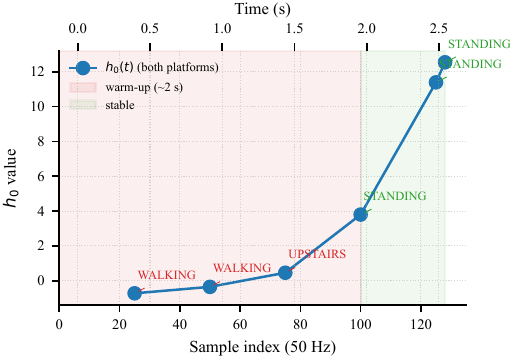}
\caption{Hidden-state component $h_0$ and emitted class label over
a single 2.56~s window (50~Hz, \textsc{standing} ground truth).
Both Arduino and MSP430 produce identical trajectories. The class
label is unstable during the recurrent warm-up; over 100 random
test windows the median stabilization point is 74 samples (1.48~s),
with a worst-case of 125 samples (2.50~s).}
\label{fig:warmup}
\end{figure}

A finding not discussed in the original \fastgrnn{}
paper~\cite{kusupati2018fastgrnn} is that prediction stability
requires a non-trivial number of samples of hidden-state
evolution before the emitted class label settles,
regardless of the underlying deployment target. Fig.~\ref{fig:warmup}
traces the emitted label and the $h_0$ component over a single
\textsc{standing} window: the model passes through \textsc{walking}
and \textsc{upstairs} predictions before locking onto the correct
class. This is a property of the recurrent
dynamics, not of either MCU; both Arduino and MSP430 produce
identical trajectories (Section~\ref{sec:results:crossplatform}).

To quantify the distribution, we evaluated 100 randomly sampled test
windows using the deployed seed-0 model and measured, for each window,
the first step $t^*$ at which the per-step prediction matches the
final-window prediction and remains stable for all subsequent steps.
The \textbf{median stabilization point is 74 samples (1.48~s)},
with an IQR of 40--86 samples (0.79--1.72~s) and a worst-case of
125 samples (2.50~s at 50~Hz).

For user-facing applications -- fall detection, gesture recognition,
posture coaching -- this implies that activity \emph{transitions}
are confirmed with a median latency of $\sim$1.5~s and a worst-case
of $\sim$2.5~s after they occur. This is a non-trivial latency budget
and should be reported alongside per-sample throughput when
characterizing on-device RNNs. We hypothesize the same effect occurs
in other gated recurrent cells of comparable size; verifying this on
LSTM/GRU baselines at matched parameter counts is an obvious
follow-up.

\subsection{Cross-Platform Determinism as Reproducibility}
\label{sec:disc:determinism}

Modern ML inference is frequently non-reproducible across hardware
due to vendor-specific BLAS, mixed-precision tensor cores, and
non-associative floating-point reductions. By contrast, our
portable C engine -- using software-emulated floating-point on
both 8-bit AVR and 16-bit MSP430 targets -- produces \emph{bit-equivalent}
predictions across platforms (Section~\ref{sec:results:crossplatform}).
This level of determinism is itself an artifact worth preserving
in safety-relevant deployments such as medical-grade HAR, where
regulatory bodies increasingly require that a software change be
re-validated only against the exact hardware on which it was
qualified. A bit-equivalent inference path across two different
ISAs is a useful escape hatch from that constraint.

\subsection{Headroom Analysis and the Pure-Q15 Dead End}
\label{sec:disc:headroom}

The \msp{} consumes 65\% of the 20~ms budget per sample, leaving
$\sim$7~ms of headroom (Section~\ref{sec:results:streaming}). In
live MPU-6050 testing this headroom comfortably absorbs an
I\textsuperscript{2}C accelerometer read, an LED status update,
and a UART log line per sample.

We did briefly explore a more aggressive \emph{pure-Q15} inference
path -- weights, activations, \emph{and} intermediate products all
in Q15 integer arithmetic. The pipeline failed empirically: the
chained per-tensor scales attenuate to $\sim$$10^{-13}$ over the
multi-stage low-rank product, which is unrepresentable in a Q15
multiplier without an int64 accumulator and a custom per-stage
shift schedule. Replacing the per-tensor scales with the simpler
fixed Q9.6 format discussed in Section~\ref{sec:method:q15}
alleviates the chaining problem but reintroduces it on the
weight side: weight magnitudes span four orders of magnitude
across tensors, so a uniform Q9.6 scale wastes up to $8\times$
of weight resolution per tensor relative to the calibrated path.
The clean fix is quantization-aware training~\cite{hubara2017qnn},
which constrains the scale schedule during training rather than
discovering it post-hoc. Given that
the LUT-based FP32 path already meets the 50~Hz real-time
constraint with comfortable headroom, we did not pursue the
QAT approach in this work; we revisit it as future work in
Section~\ref{sec:disc:future}.

\subsection{Limitations}
\label{sec:disc:limitations}

Three limitations bound the scope of our claims:

\begin{itemize}
\item \textbf{Dataset.} \hapt{} is recorded under laboratory
conditions with a fixed waist-mounted smartphone. On-body sensor
displacement, free-living data distribution shift, and
inter-subject variation are known to degrade HAR
accuracy~\cite{wang2019harsurvey,demrozi2020harsurvey} and are not
modelled here.

\item \textbf{Label set.} We train and evaluate on the six basic
\hapt{} activities. The full \hapt{} label set adds six postural
transitions (e.g.\ \textsc{stand-to-sit}, \textsc{lie-to-sit})
which our model does not handle. This is a deliberate choice
to match the original \fastgrnn{} HAR benchmark.

\item \textbf{The DOWNSTAIRS class.} \textsc{downstairs} is the
binding-constraint class throughout the pipeline
(Fig.~\ref{fig:perclass}), and we did not attempt class-specific
remediation in this paper. Section~\ref{sec:disc:future} sketches
several plausible interventions.

\item \textbf{Statistical significance testing.} Our multi-seed
evaluation reports mean$\pm$standard deviation across five seeds
but does not include paired significance tests or non-parametric
bootstrap intervals. Formal significance testing is left for
future work.
\end{itemize}

\subsection{Future Directions}
\label{sec:disc:future}

Four extensions are natural next steps from this work.

\begin{itemize}
\item \textbf{Dual-rank static-vs-dynamic decomposition.} In our
rank sweep (Section~\ref{sec:results:lowrank}), $r_u{=}4$ is
favorable for the dynamic classes (\textsc{walking},
\textsc{upstairs}, \textsc{downstairs}) and $r_u{=}8$ is favorable
for the static classes (\textsc{sitting}, \textsc{standing}).
A simple low-cost variant -- $\mathbf{U}_\text{eff} = \mathrm{LowRank}(r{=}4) + \mathrm{diag}(\boldsymbol{\alpha})$
-- adds only $H$ extra parameters while letting a static
DC-like signal pass through the diagonal residual and a dynamic
pattern through the low-rank branch. We expect this to recover
the dynamic-class accuracy of $r_u{=}4$ without the static-class
degradation.

\item \textbf{DOWNSTAIRS-targeted features.} The standard
UCI~HAR pipeline applies a Butterworth low-pass at 0.3~Hz to
separate gravity from body acceleration; we omitted this step to
match the original \fastgrnn{} HAR setup. Re-introducing it, or
augmenting the input with a small number of FFT-band features over
the 128-sample window, is the lowest-cost path to closing the gap
on the \textsc{downstairs} class.

\item \textbf{Quantization-aware training for pure-Q15 inference.}
As discussed in Section~\ref{sec:disc:headroom}, post-training
quantization fails for a fully integer pipeline because of scale
chaining; QAT~\cite{hubara2017qnn} should resolve this. A pure-Q15
path would likely yield an additional 2--3$\times$ speedup on the
\msp{} and is the natural successor to the LUT trick.

\item \textbf{Transition states.} Adding the six postural transitions
of the full \hapt{} label set raises the practical question of
how to allocate the recurrent state between long-horizon
activity recognition and short-horizon transition detection.
A multi-scale recurrent cell, evaluated at two cadences, is a
straightforward starting point.
\end{itemize}

\section{Conclusion}
\label{sec:conclusion}

We reproduced \fastgrnn{} end-to-end on two bare-metal
microcontrollers -- the 8-bit \arduino{} and the 16-bit
multiplier-less \msp{} -- and demonstrated that compact gated
recurrent cells are a practical fit for kilobyte-class wearable
HAR. The deployed model occupies 566~bytes of weights, achieves
macro F1~$=$~0.918 (seed~0) on the \hapt{} test set --- with a
five-seed Q15 mean of $0.853{\pm}0.107$ --- and runs in real time
at 50~Hz on both targets with zero over-budget samples. The
integer C inference engine is bit-equivalent across 8-bit AVR and
16-bit MSP430 implementations.

Four contributions extend the original \fastgrnn{} paper:
cross-platform deterministic inference, a 30.5$\times$ LUT-based
speedup for multiplier-less MCUs, empirical characterization of the
recurrent warm-up latency (median 74 samples / 1.48~s; worst-case
2.50~s), and a hardware energy characterization showing 17.7~mW
active power, $<$0.09~mW idle power, and a 96.7\% reduction in
per-window energy enabled by the LUT. A 566-byte model running at
50~Hz with a coin-cell power budget is a concrete existence proof
that calibrated quantization, look-up-table activations, and
measured energy profiling can bring recurrent HAR to the silicon we
already have. All code, models, and reproducibility artifacts are
publicly available at \url{https://github.com/emre1998/fastgrnn-har}
under the Apache License 2.0.

\bibliographystyle{IEEEtran}
\bibliography{references}

@inproceedings{kusupati2018fastgrnn,
  title     = {{FastGRNN}: A Fast, Accurate, Stable and Tiny Kilobyte
               Sized Gated Recurrent Neural Network},
  author    = {Kusupati, Aditya and Singh, Manish and Bhatia, Kush and
               Kumar, Ashish and Jain, Prateek and Varma, Manik},
  booktitle = {Advances in Neural Information Processing Systems (NeurIPS)},
  year      = {2018}
}

@inproceedings{kumar2017bonsai,
  title     = {Resource-Efficient Machine Learning in 2~{KB} {RAM} for
               the Internet of Things},
  author    = {Kumar, Ashish and Goyal, Saurabh and Varma, Manik},
  booktitle = {International Conference on Machine Learning (ICML)},
  year      = {2017}
}

@inproceedings{gupta2017protonn,
  title     = {{ProtoNN}: Compressed and Accurate {k}{NN} for
               Resource-Scarce Devices},
  author    = {Gupta, Chirag and Suggala, Arun Sai and Goyal, Ankit and
               Simhadri, Harsha Vardhan and Paranjape, Bhargavi and
               Kumar, Ashish and Goyal, Saurabh and Udupa, Raghavendra
               and Varma, Manik and Jain, Prateek},
  booktitle = {International Conference on Machine Learning (ICML)},
  year      = {2017}
}

@misc{edgeml,
  author = {{Microsoft Research India}},
  title  = {{EdgeML}: Machine learning for resource-constrained edge devices},
  year   = {2017--2024},
  note   = {\url{https://github.com/microsoft/EdgeML}}
}

@inproceedings{anguita2013public,
  title     = {A Public Domain Dataset for Human Activity Recognition
               Using Smartphones},
  author    = {Anguita, Davide and Ghio, Alessandro and Oneto, Luca and
               Parra, Xavier and Reyes-Ortiz, Jorge L.},
  booktitle = {European Symposium on Artificial Neural Networks (ESANN)},
  year      = {2013}
}

@article{reyes2015transition,
  title   = {Transition-Aware Human Activity Recognition Using Smartphones},
  author  = {Reyes-Ortiz, Jorge-L. and Oneto, Luca and Sam{\`a}, Albert
             and Parra, Xavier and Anguita, Davide},
  journal = {Neurocomputing},
  year    = {2016},
  volume  = {171},
  pages   = {754--767}
}

@article{hochreiter1997lstm,
  title   = {Long Short-Term Memory},
  author  = {Hochreiter, Sepp and Schmidhuber, J{\"u}rgen},
  journal = {Neural Computation},
  volume  = {9},
  number  = {8},
  pages   = {1735--1780},
  year    = {1997}
}

@inproceedings{cho2014gru,
  title     = {Learning Phrase Representations Using {RNN}
               Encoder-Decoder for Statistical Machine Translation},
  author    = {Cho, Kyunghyun and van Merri{\"e}nboer, Bart and Gulcehre,
               Caglar and Bahdanau, Dzmitry and Bougares, Fethi and
               Schwenk, Holger and Bengio, Yoshua},
  booktitle = {Conference on Empirical Methods in Natural Language
               Processing (EMNLP)},
  year      = {2014}
}

@inproceedings{jacob2018integer,
  title     = {Quantization and Training of Neural Networks for Efficient
               Integer-Arithmetic-Only Inference},
  author    = {Jacob, Benoit and Kligys, Skirmantas and Chen, Bo and Zhu,
               Menglong and Tang, Matthew and Howard, Andrew and Adam,
               Hartwig and Kalenichenko, Dmitry},
  booktitle = {IEEE Conference on Computer Vision and Pattern Recognition
               (CVPR)},
  year      = {2018}
}

@inproceedings{han2016deepcompression,
  title     = {Deep Compression: Compressing Deep Neural Networks with
               Pruning, Trained Quantization and {Huffman} Coding},
  author    = {Han, Song and Mao, Huizi and Dally, William J.},
  booktitle = {International Conference on Learning Representations (ICLR)},
  year      = {2016}
}

@inproceedings{han2015pruning,
  title     = {Learning Both Weights and Connections for Efficient Neural
               Networks},
  author    = {Han, Song and Pool, Jeff and Tran, John and Dally, William J.},
  booktitle = {Advances in Neural Information Processing Systems (NeurIPS)},
  year      = {2015}
}

@article{hubara2017qnn,
  title   = {Quantized Neural Networks: Training Neural Networks with
             Low Precision Weights and Activations},
  author  = {Hubara, Itay and Courbariaux, Matthieu and Soudry, Daniel
             and El-Yaniv, Ran and Bengio, Yoshua},
  journal = {Journal of Machine Learning Research},
  volume  = {18},
  pages   = {1--30},
  year    = {2017}
}

@article{blumensath2009iht,
  title   = {Iterative Hard Thresholding for Compressed Sensing},
  author  = {Blumensath, Thomas and Davies, Mike E.},
  journal = {Applied and Computational Harmonic Analysis},
  volume  = {27},
  number  = {3},
  pages   = {265--274},
  year    = {2009}
}

@inproceedings{frankle2019lt,
  title     = {The Lottery Ticket Hypothesis: Finding Sparse, Trainable
               Neural Networks},
  author    = {Frankle, Jonathan and Carbin, Michael},
  booktitle = {International Conference on Learning Representations (ICLR)},
  year      = {2019}
}

@inproceedings{banbury2021mlperf,
  title     = {{MLPerf} Tiny Benchmark},
  author    = {Banbury, Colby and Reddi, Vijay Janapa and Torelli, Peter
               and Holleman, Jeremy and Jeffries, Nat and Kiraly, Csaba
               and Montino, Pietro and Kanter, David and Ahmed, Sebastian
               and Pau, Danilo and others},
  booktitle = {Conference on Machine Learning and Systems (MLSys)},
  year      = {2021}
}

@inproceedings{lin2020mcunet,
  title     = {{MCUNet}: Tiny Deep Learning on {IoT} Devices},
  author    = {Lin, Ji and Chen, Wei-Ming and Lin, Yujun and Cohn, John
               and Gan, Chuang and Han, Song},
  booktitle = {Advances in Neural Information Processing Systems (NeurIPS)},
  year      = {2020}
}

@article{capotondi2020cmixnn,
  title   = {{CMix-NN}: Mixed Low-Precision {CNN} Library for
             Memory-Constrained Edge Devices},
  author  = {Capotondi, Alessandro and Rusci, Manuele and Fariselli,
             Marco and Benini, Luca},
  journal = {IEEE Transactions on Circuits and Systems II},
  year    = {2020}
}

@article{david2021tflitemicro,
  title   = {{TensorFlow Lite Micro}: Embedded Machine Learning for
             {TinyML} Systems},
  author  = {David, Robert and Duke, Jared and Jain, Advait and Janapa
             Reddi, Vijay and Jeffries, Nat and Li, Jian and Kreeger,
             Nick and Nappier, Ian and Natraj, Meghna and Regev, Shlomi
             and others},
  journal = {Conference on Machine Learning and Systems (MLSys)},
  year    = {2021}
}

@inproceedings{lai2018cmsisnn,
  title     = {{CMSIS-NN}: Efficient Neural Network Kernels for {ARM}
               {Cortex-M} {CPUs}},
  author    = {Lai, Liangzhen and Suda, Naveen and Chandra, Vikas},
  booktitle = {arXiv preprint arXiv:1801.06601},
  year      = {2018}
}

@article{ordonez2016deepconvlstm,
  title   = {Deep Convolutional and {LSTM} Recurrent Neural Networks for
             Multimodal Wearable Activity Recognition},
  author  = {Ord{\'o}{\~n}ez, Francisco Javier and Roggen, Daniel},
  journal = {Sensors},
  volume  = {16},
  number  = {1},
  year    = {2016}
}

@article{wang2019harsurvey,
  title   = {Deep Learning for Sensor-Based Activity Recognition: A
             Survey},
  author  = {Wang, Jindong and Chen, Yiqiang and Hao, Shuji and Peng,
             Xiaohui and Hu, Lisha},
  journal = {Pattern Recognition Letters},
  volume  = {119},
  pages   = {3--11},
  year    = {2019}
}

@article{demrozi2020harsurvey,
  title   = {Human Activity Recognition Using Inertial, Physiological
             and Environmental Sensors: A Comprehensive Survey},
  author  = {Demrozi, Florenc and Pravadelli, Graziano and Bihorac,
             Azra and Rashidi, Parisa},
  journal = {IEEE Access},
  volume  = {8},
  pages   = {210816--210836},
  year    = {2020}
}

@inproceedings{strubell2019energy,
  title     = {Energy and Policy Considerations for Deep Learning in {NLP}},
  author    = {Strubell, Emma and Ganesh, Ananya and McCallum, Andrew},
  booktitle = {Annual Meeting of the Association for Computational
               Linguistics (ACL)},
  year      = {2019}
}

@article{patterson2021carbon,
  title   = {Carbon Emissions and Large Neural Network Training},
  author  = {Patterson, David and Gonzalez, Joseph and Le, Quoc and
             Liang, Chen and Munguia, Lluis-Miquel and Rothchild,
             Daniel and So, David and Texier, Maud and Dean, Jeff},
  journal = {arXiv preprint arXiv:2104.10350},
  year    = {2021}
}

@misc{burkacky2022chipshortage,
  author = {Burkacky, Ondrej and Lingemann, Stephanie and Pototzky,
            Katja},
  title  = {Coping with the Auto-Semiconductor Shortage: Strategies for
            Success},
  howpublished = {McKinsey \& Company},
  year   = {2022},
  note   = {\url{https://www.mckinsey.com/industries/automotive-and-assembly/our-insights}}
}

@techreport{khan2021chipshortage,
  author      = {Khan, Saif M. and Mann, Alexander and Peterson, Dahlia},
  title       = {The Semiconductor Supply Chain: Assessing National
                 Competitiveness},
  institution = {Center for Security and Emerging Technology (CSET),
                 Georgetown University},
  year        = {2021}
}

@misc{ti_msp430,
  author = {{Texas Instruments}},
  title  = {{MSP430x2xx} Family User's Guide (SLAU144J)},
  year   = {2013},
  note   = {\url{https://www.ti.com/lit/ug/slau144j/slau144j.pdf}}
}

\appendices
\section{Hyperparameters}
\label{app:hyper}

Table~\ref{tab:hyper} lists the full hyperparameter set used to
produce the deployed model. All values are unchanged across the
five training seeds.

\begin{table}[h]
\centering
\caption{Training and deployment hyperparameters for the deployed
\fastgrnn{} model.}
\label{tab:hyper}
\resizebox{\columnwidth}{!}{%
\begin{tabular}{lr}
\toprule
Parameter & Value \\
\midrule
Hidden size $H$                  & 16 \\
Input dimensionality $d$         & 3 (tri-axial accel.) \\
Window length $T$                & 128 samples (2.56~s) \\
Sample rate                      & 50~Hz \\
Output classes                   & 6 \\
\midrule
Recurrent rank $r_u$             & 8 \\
Input rank $r_w$                 & 2 \\
Target sparsity $s$              & 0.5 (181 cell $+$ 102 head $=$ 283 nonzero / 566~B) \\
Sparsity ramp epochs             & 50 (cubic schedule) \\
Frozen-mask fine-tune epochs     & 50 \\
\midrule
Optimizer                        & Adam \\
Learning rate $\eta$             & $10^{-3}$ \\
Batch size                       & 64 \\
Total epochs                     & 100 \\
Random seeds                     & $\{0, 1, 2, 3, 4\}$ \\
\midrule
Activation function              & sigmoid, tanh \\
LUT size                         & 256 entries each \\
LUT input domain                 & $[-8, +8]$ \\
Weight quantization              & per-tensor Q15 \\
Activation quantization          & per-tensor Q15 + calibration \\
Calibration mini-batches         & 5 \\
Calibration headroom             & 10\% \\
\bottomrule
\end{tabular}}
\end{table}

\section{Q15 Quantization Implementation}
\label{app:q15}

The per-tensor Q15 quantization step used in
Section~\ref{sec:method:q15} is a one-line operation per tensor. In
Python:

\begin{verbatim}
def to_q15(W: np.ndarray, scale: float) -> np.ndarray:
    return (W / scale).round() \
                      .clip(-32768, 32767) \
                      .astype(np.int16)
\end{verbatim}

The per-tensor scale is chosen as
\[
s_\ell = \max_{ij}\bigl|W^{(\ell)}_{ij}\bigr| \,\big/\, 32767,
\]
which is the largest scale that keeps every quantized entry inside
the signed 16-bit range. The same expression is used for both the
recurrent weight factors $\mathbf{W}_1, \mathbf{W}_2,
\mathbf{U}_1, \mathbf{U}_2$ and the classifier head. At runtime, the
dequantized value is computed as

\begin{verbatim}
float w = (float)W_q15[i] * scale;
\end{verbatim}

The runtime cost per dequantize is one int-to-float conversion and
one multiply; both are cheap on the AVR and, with the LUT
intercepting the activations, manageable on the multiplier-less
\msp{}.

\section{LUT Generation Algorithm}
\label{app:lut}

The sigmoid and tanh look-up tables of Section~\ref{sec:method:lut}
are pre-computed offline in Python and emitted as a C header:

\begin{verbatim}
LUT_SIZE     = 256
INPUT_MIN    = -8.0
INPUT_MAX    =  8.0
BUCKET_WIDTH = (INPUT_MAX - INPUT_MIN) / LUT_SIZE

sigmoid_lut = [
    1.0 / (1.0 + math.exp(-(INPUT_MIN
            + (i + 0.5) * BUCKET_WIDTH)))
    for i in range(LUT_SIZE)
]
tanh_lut = [
    math.tanh(INPUT_MIN + (i + 0.5) * BUCKET_WIDTH)
    for i in range(LUT_SIZE)
]
\end{verbatim}

Each entry stores the value of the activation at the
\emph{center} of its bucket (the $(i + 0.5)$ offset), which is
the maximum-likelihood estimate for a uniformly distributed
sub-bucket input and avoids a half-bucket bias. The runtime lookup is

\begin{verbatim}
static inline float lut_eval(const float *lut, float x) {
    if (x <= LUT_INPUT_MIN) return lut[0];
    if (x >= LUT_INPUT_MAX) return lut[LUT_SIZE - 1];
    int idx = (int)((x - LUT_INPUT_MIN) * LUT_INPUT_SCALE);
    return lut[idx];
}
\end{verbatim}

with \texttt{LUT\_INPUT\_SCALE = 1 / BUCKET\_WIDTH}. The two tables
together occupy 2~KB of Flash; together they eliminate every
\texttt{expf} and \texttt{tanhf} call from the runtime.

\section{MSP430 I\textsuperscript{2}C Bring-Up}
\label{app:hw}

Live MPU-6050 read-out on the \msp{} uses the USCI\_B0 peripheral
in I\textsuperscript{2}C master mode with the standard pin
assignment (P1.6~$=$~SCL, P1.7~$=$~SDA). Two practical issues are
worth recording for future readers:

\begin{itemize}
\item \textbf{J5 jumper removal.} The MSP-EXP430G2 LaunchPad
multiplexes the on-board green LED with P1.6. The J5 jumper must
be physically removed before any I\textsuperscript{2}C
communication; otherwise SCL is loaded by the LED and the bus
sits below the I\textsuperscript{2}C high-level threshold.

\item \textbf{Bus-busy recovery.} The USCI\_B0 peripheral can latch
into a \texttt{UCSCLLOW + UCBBUSY} state if power is removed
mid-transfer; the conventional 9-clock recovery sequence
(toggle SCL nine times with SDA released, then STOP) restores the
bus. A GPIO-level health check at boot
(\texttt{P1IN \& (BIT6|BIT7)} should be \texttt{0xC0}) detects the
condition before the I\textsuperscript{2}C state machine is engaged.
\end{itemize}

These two items consumed approximately one day of bring-up time
and are not documented in the standard MSP430 application
notes~\cite{ti_msp430}. They are included here in case they save
the next reader the same day.

\end{document}